# Distinct moiré Exciton dynamics in $WS_2$/ $WSe_2$ heterostructure

Feng Kai,[1] Xiong Wang,[1] Yiqin Xie,[1] Yuhui Yang,[1] Kenji Watanabe,[4] Takashi Taniguchi,[4] Hongyi Yu,[2,3] Wang Yao,[1] Xiaodong Cui[1]

[1]Physics Department and HK Institute of Quantum Science & Technology, The University of Hong Kong, Pokfulam road, Hong Kong, China

[2]Guangdong Provincial Key Laboratory of Quantum Metrology and Sensing & School of Physics and Astronomy, Sun Yat-sen University, Zhuhai Campus, Zhuhai, 519082, China

[3]State Key Laboratory of Optoelectronic Materials and Technologies, Sun Yat-sen University, Guangzhou Campus, Guangzhou, 510275, China

[4]Advanced Materials Laboratory, National Institute for Materials Science, 1-1 Namiki, Tsukuba, 305-0044, Japan.

## Abstract

This letter reports a time resolved pump-probe reflectance spectroscopic study on moiré excitons in a twisted monolayer $WS_2/WSe_2$ heterostructure. By probing at the resonant energies of intralayer excitons, we observed their distinct temporal tracks under the influence of interlayer excitons, which we attribute to the discrepancy in spatial distribution of the intralayer excitons in different layers. We also observed that intralayer moiré excitons in $WSe_2$ layer differ at decay rate, which reflects different locations of Wannier-like and charge-transfer intralayer excitons in a moiré cell. We concluded that the interlayer moiré excitons form within a few picoseconds and have the lifetime exceeding five nanoseconds. Our results provide insights into the nature of moiré excitons and the strain's significant impact on their behaviour in twisted heterostructures, which could have important implications for the development of novel optoelectronic devices.

## Introduction

The emergent van der Waal (vdW) heterostructures provide a platform that allows manipulation of various degrees of freedoms and creation of artificial lattices. The deliberate lattice mismatch between individual vdW layers engenders a periodic interference pattern, known as the moiré pattern. This tailored moiré pattern offers a great flexibility of building artificial superlattices and has invoked fantastic quantum states including superconducting, fractional quantum anomalous Hall, strongly correlated states, etc[1-11]. One of the intriguing topics is the exciton physics in moiré superlattices, particular TMD moiré superlattices which feature direct bandgap in monolayers and rich degrees of freedoms. Predominantly, TMD heterostructures display a type-II quantum well band alignment, with conduction and valence band edges residing in disparate layers. Consequently, photocarriers, upon optical excitation, swiftly relax to the disparate band edges through interlayer charge transfer, culminating in the formation of interlayer excitons. The spatial separation in these structures renders the interlayer excitons an extended lifetime, often reaching several nanoseconds[12, 13] and ultralong valley lifetime around *40us*[14, 15]. This long-life significantly lowers the barriers towards quantum phase transitions. In moiré superlattices these interlayer excitons modulated by the periodic moiré potentials display exotic features originating from the unique dispersion in the superlattice. Besides the moiré potential from the heterostructure stacking, the in-plane strain in individual layer arising from the lattice mismatch also contribute to the moiré potential modulation[16-24]. Given the disparity in Young's modulus and fracture strength[25], the strain within each heterostructure layer varies in intensity. For interlayer excitons, this results in electrons and holes in opposing layers experiencing distinct in-plane moiré potential modulations. These multifaceted moiré potentials complicate the microscopic understanding of moiré excitons. As yet, the dynamics of moiré excitons remain largely unexplored.

The lack of dynamic study of the interlayer moiré excitons partially results from the relatively weak oscillator strength of the interlayer excitons. The spatial separation reduces the electron-hole wave function overlap and weakens the oscillator strength of interlayer excitons. Consequently, interlayer exciton is usually invisible in the reflection/absorption spectra which are the most popular techniques in ultrafast dynamics study. In this letter, we report a time-resolved pump-probe reflection spectroscopic study on moiré excitons of the twisted monolayer $WSe_2/WS_2$ heterostructure. We probed the reflectance spectra at the resonant energies of $WSe_2$ and $WS_2$ intralayer excitons and observed the different carrier dynamic patterns. These different dynamics reflect the coupling between intralayer excitons and interlayer excitons in a moiré cell, respectively. Our findings reveal a significant in-plane moiré strain in $WSe_2$ layer and the difference in $WSe_2$ and $WS_2$ intralayer excitons under the strain modulation.

## Fabrication

Monolayer $WSe_2$ and $WS_2$ were mechanically exfoliated from single crystal and were vertically stacked as a heterostructure with a twist angle of around 0.8°. The heterostructure was encapsulated single crystal hexagonal boron nitride (h-BN) nanoflakes, then the whole stack was transferred onto a silicon substrate with 90nm silicon dioxide cap-layer. The twist angle between the monolayer $WS_2$ and $WSe_2$ was identified by the polarization-dependent second harmonic generation (SHG) measurement (Data shown in Figure 1). The two slightly mis-aligned six-fold SHG patterns were clearly observed (three folds is shown here), and the six-fold fitting revealed the relative twist angle of 0.8°. The heterostructure yields a stronger SHG intensity (Figure 1b) than both individual monolayer $WSe_2$ and $WS_2$, indicating the heterostructure has AA stacking configuration (R-stacking).

## Interlayer exciton

Figure 1c shows the photoluminescence spectra collected on the heterostructure and monolayer $WSe_2$ at the base temperature of *20K*. The photoluminescence peaks(red) of neutral excitons at *1.734eV* and trions (charged exciton) at1.698eV of monolayer $WSe_2$ fade in the heterostructure as a consequence of interlayer charge transfer. This is consistent with the type-II band alignment (Figure 1d) in which the minimum of conduction band is located at K(K') valley of $WS_2$ and the maximum of valence band at $WSe_2$ layer. Under optical excitation, most intralayer electron-hole pairs are disassociated and drift to opposite layers, namely, holes transferring to $WSe_2$ while electrons to $WS_2$. This interlayer charge transfer takes place in around hundreds of femtoseconds[26-30] which is a few orders of magnitude faster than the intralayer exciton lifetime, consequently, quenches the PL of intralayer excitons. The band edge electron-hole pairs at different layers are bonded to form interlayer excitons, identified as the peaks around 1.395eV (Figure 1c, black).

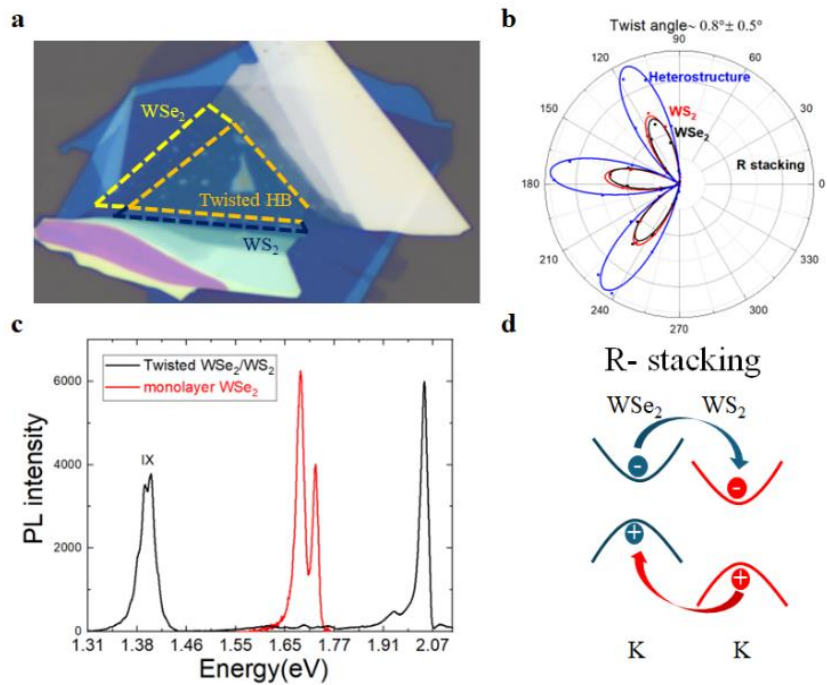

**Fig. 1 | a,** Optical microscopic image of the device. Heterostructure, monolayer $WSe_2$ and $WS_2$ region are outlined by orange, yellow and black dash line, respectively. **b,** Polarization-resolved SHG of individual monolayers $WSe_2$ (black), $WS_2$ (red) and heterostructure(blue). The relative twist angle is extracted to be 0.8°±0.5° by fitting. **c,** Photoluminescence of monolayer $WSe_2$ (red) and $WSe_2/WS_2$ heterostructure device(black) under 2.087eV(594nm) excitation at intensity of *5uW*. The interlayer exciton peak appears at *1.395eV*, labelled as *IX*. **d,** Diagram of the electron and hole transfer in type-II band alignment.

## Ladder-like interlayer exciton emission

To further investigate the interlayer exciton in our device, we did a series of intensity-dependent photoluminescence measurements. Figure 2a presents photoluminescence of interlayer excitons at different excitation intensities. Due to the interference of our CCD device at the specific wavelength, the raw data (grey lines) carries significant interference patterns. To remove artificial device noise, we use multi-peak Lorentz fitting to extract the PL spectra. We observed that there's only one interlayer exciton emission peak ($IX_1$) at *1.388eV* under low excitation intensity of *0.5uW* (Figure 2a, black line). As the excitation intensity increases to *50uW*, the second interlayer emission peak ($IX_2$) arises at the energy *37meV* higher than $IX_1$(blue line). Under low excitation intensity, the interlayer excitons density, or the filling factor (the number of excitons per moiré unit cell), is low, and the exciton-exciton interaction is negligible. As the excitation intensity increases, the interlayer excitons populates and the exciton-exciton interaction arises. When the filling factor is above 1, namely, every moiré cell has one exciton and some moiré cells has to accommodate two excitons, the exciton energy is lifted owing to the onsite dipole-dipole repulsion. This is well described by the exciton Hubbard model[31]. Our photoluminescence experiment indicates an onsite dipole-dipole repulsion energy($U$) of *37meV*, which manifests as the energy jump from $IX_1$ to $IX_2$ (Figure 2b), qualitatively consistent with the previous reports[31]. With such understanding, the thirdly and fourthly occupied moiré cell correspond to $IX_3$ and $IX_4$, with the energy jump of $\Delta E_{32}=29meV$ and $\Delta E_{43}=49meV$, shown in Figure 2b. Besides the energy jump between *IXs*, all the four peaks blue-shift with the increasing excitation intensity from *0.5* to *1088uW*. The blue shift originates from long-range dipolar repulsion between interlayer excitons trapped in the neighboring moiré cell[31].

## Reflection contrast

We then characterized the absorption by static reflection contrast spectroscopy. Figure 2c shows the background-free reflection contrast (RC) of the heterostructure device and monolayers. Here, the RC is obtained by $R/R_0-R_{bg}$, where $R$ and $R_0$ are the reflection spectrum of the heterostructure device and its hBN-only region, and $R_{bg}$ is the background in RC due to substrate interference. We found the *A* exciton absorption peak (*1.727eV*, red line) of monolayer $WSe_2$ splits into three peaks (with the energies of *1.683eV*, *1.742eV* and *1.784eV*, labelled as *I, II, III*, respectively) in the heterostructure. These emergent three peaks were previously observed and attributed to the unique moiré exciton feature in $WSe_2/WS_2$ heterostructure[32]. It was interpreted with the continuum model that the emergent periodic moiré potential modifies the excitonic dispersion, splitting the edge into three sub bands, contributing to the observed three peaks[32-34].

However, we also note that the $WS_2$ layer, sharing the moiré pattern with $WSe_2$, does NOT experience splitting. This could be attributed to the different in-plane strain patterns in the two layers under the lattice reconstructions [23]. It has been shown that, in a rigid moiré landscape without the lattice reconstruction, the moiré potential experienced by intralayer excitons in either layer is very weak[35]. When structure relaxation is taken into account, the resultant periodic strains can give rise to pronounced modulation of electronic band structures and the energy landscape of the intralayer exciton [36]. Driven by the stacking-dependent interlayer van der Waals interaction, the $WS_2/WSe_2$ heterostructure, once fabricated, would experiences a structural reconstruction[23]. Given the $WS_2$ has higher Young's modulus(*302GPa*) and fracture strength(*47GPa*) than $WSe_2$(*258GPa and 38GPa*)[25], $WSe_2$ and $WS_2$ experience different extent of lattice reconstruction and the in-plane strain modification. Our result suggests that a stronger strain pattern is generated in the $WSe_2$ layer upon the lattice reconstruction, giving rise to the strong moiré potential and miniband splitting for $WSe_2$ exciton, while the strain in $WS_2$ layer is not strong enough to split its intralayer exciton dispersion. We further infer that the $WS_2$ intralayer exciton remains largely similar to that in pristine $WS_2$ monolayer due to its strong mechanical strength.

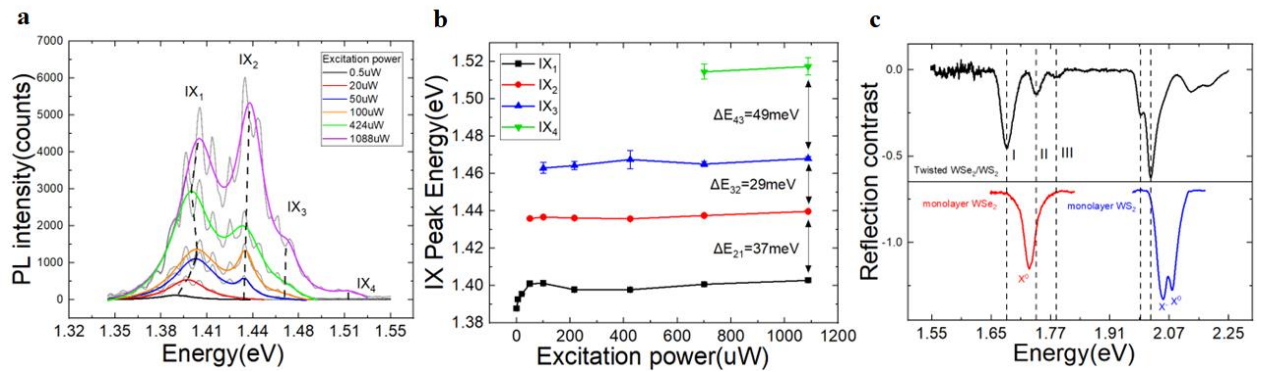

Figure 2 | Photoluminescence and reflection contrast spectra of WSe$_2$/WS$_2$ moiré superlattice at *20K*.

**a,** Excitation intensity-dependent interlayer exciton photoluminescence: raw data (grey lines) and the corresponding Lorentz fitting curves (colour lines). The obtained four interlayer exciton peaks, labelled as *IX$_1$* to *IX$_4$*, are outlined by dashed black lines, respectively. **b,** Extracted *IX* peak energies vs excitation intensity from Figure 2a. **c,** Background-free reflection contrast (RC) of the WSe$_2$/ WS$_2$ device (top, black). In the bottom, RC of monolayer WSe$_2$ (red) and WS$_2$ (blue). The intrinsic WSe$_2$ *A* exciton at *1.727eV*(718nm) splits into three moiré exciton resonances in the heterostructure, labelled as *I, II, III*. Three resonances have energies of *1.784eV(695nm)*, *1.742eV(712nm)*, *1.683eV(737nm)*, respectively.

## Pump probe spectroscopy on moiré superlattice

After characterising the effects of moiré landscape on the intralayer excitons in our device from the photoluminescence and reflection contrast spectra, we investigate the moiré exciton dynamics by pump probe spectroscopy. Since the interlayer exciton has very weak oscillator strength, the pump probe spectroscopy that based on reflection change can't capture it directly. Instead, we probe around the resonant energies of the intralayer excitons of WSe$_2$ and WS$_2$. Here we pump the heterostructure with a beam of *3.1eV(400nm)* at *100uW* that is energetic enough to excite all low-energy transitions. The hot photo carriers, electron-hole pairs, thermally relax in the first hundreds of femtosecond[37, 38], followed by charge transfer and interlayer exciton formation in around 1*ps*[39]. Owing to the type-*II* band alignment in WSe$_2$/ WS$_2$, the *IX*'s electrons (holes) reside in the conduction band of WS$_2$ layer (valence band of WSe$_2$ layer), which share the same band as the intralayer excitons of individual layers. By probing at the intralayer excitons, we can monitor the dynamics of the electron or hole components of moiré interlayer excitons, namely, we monitor the impact of pump-induced interlayer excitons on the intralayer exciton transitions

## Probe around resonance of WS$_2$ intralayer excitons

Figure 3a shows the pump-induced reflection change(*dR/R*) as a function of the delay time and the probe energy across the range of the WS$_2$'s intralayer exciton transition. The transient reflectance of *dR/R* > 0 represented by bright-colored region implies that the sample reflectance increases under the increased pump excitation intensity, and vice versa. To analyze the *dR/R* map, we horizontally slice the *dR/R* map at different delay time (Figure 3b) to get the profile of *dR/R vs* energy. We notice that the *dR/R* prior to the pump-probe coincidence *t=0* shows a non-zero value, significantly higher than the background noise. As the laser pulse repeats with a time interval of 13.2*ns(76MHz)*, the non-zero *dR/R* at *t<0* indicates that a significant portion of interlayer excitons survive longer than 13.2*ns*. Since the profile of *dR/R vs* energy is influenced by the resonant energy shift and the suppression of neutral exciton($X^0$) and trion($X^-$) oscillator strength owing to state filling. To quantitively extract the energy shift, we append the pump induced reflection change (*dR*) to the static reflection contrast(*R/R$_0$*), constructing the dynamic reflection contrast at various time (Figure 3c). With Lorentz fitting, we extract the evolution of the energy shifts for the absorption peaks (Figure 3d). The energy shifts of both resonances show flat decay pattern in the probe time ranging from *t= 20ps* to *80ps*(Figure 3d), which is another evidence for interlayer excitons' long lifetime In contrast, the exciton lifetime in monolayer $WS_2$ determined by the pump-probe spectroscopy is around $10^2 ps$ [40]. This long lifetime probed at the resonant energy of *A* exciton doesn't reflect the dynamics of intralayer excitons in $WS_2$ layer. Instead, this probe at intralayer exciton energy reflects the lifetime of the interlayer excitons as the interlayer excitons and delocalized intralayer excitons of $WS_2$ layer share the same electron orbits. This provides an easy approach to address the dynamics of interlayer exciton with pump-probe spectroscopy.

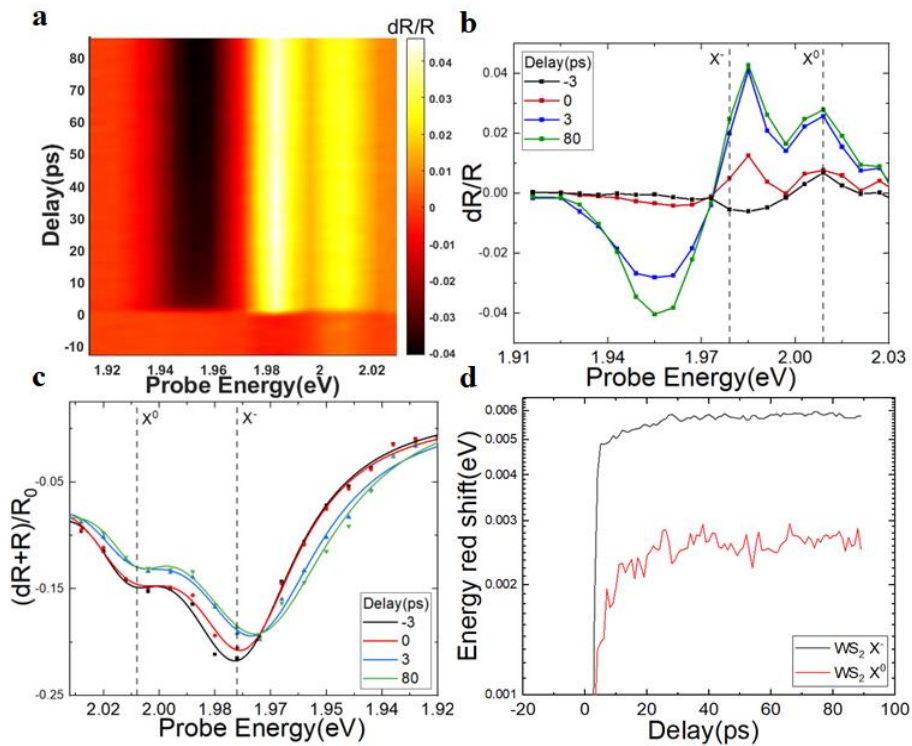

**Figure 3 | Pump probe spectroscopy for $WS_2$ intralayer transition**

**a,** Pump-induced reflection change map*(dR/R)* as a function of delay and probe energy, where *R* is the reflection spectra of twisted $WSe_2/WS_2$ heterostructure. **b,** *dR/R* vs Probe energy at *-3, 0, 3, 80ps*. The neutral exciton ($X^0$) and Trion($X^-$) is labelled with dash line for reference. **c,** Reflection contrast at different delays. **d,** Energy red shift vs delays for the neutral exciton($X^0$) and Trion($X^-$).

## Probe around the resonance energy of $WSe_2$ intralayer excitons

We probed the $WSe_2$'s intralayer exciton transition to investigate the influence of moiré interlayer excitons on $WSe_2$. We observed a negative *dR/R* region from *1.65* to *1.68eV* (Figure 4a, dark region) accompanied by a strong positive region from *1.68* to *1.72eV* (Figure 4a, bright region). This can be explained by the red shift and suppression of moiré resonance *I*, as diagramed in Figure 4c. Another positive region appears in the range from *1.73* to *1.8eV*, covering both resonance II and III. To quantitively extract the energy shift and absorption suppression, we use the same method as we did in $WS_2$ to get the reflection contrast at various delays, $(dR+R)/R_0$, as presented in Figure 4d, and then extract the energy shifts and the amplitude suppression with Lorentz line shape fitting.

We found that the peak *I* and *II* red shift, and in contrast, the peak III blue shifts upon pump (Figure 4e). Here we consider several possible mechanisms for the energy shift: (1) The bandgap renormalization[41, 42] caused by Coulomb screening of free carriers or excitons on Coulomb repulsion in lattice, resulting in electronic bandgap shrink. (2) The screening effect[43-45] caused by exciton or free carrier's Coulomb screening on exciton which reduces the exciton binding energy and makes the exciton resonant energy blue shift. Since the moiré exciton *III* has larger Bohr radius(around *5nm*) than moiré exciton *I* (tightly bound)[36], exciton *III* is more sensitive to the dielectric environment and consequently experiences stronger charge-exciton and exciton-exciton screening effect. The significantly higher Coulomb screening on large Bohr radius excitons might be the reason for the blue shift. The sensitivity of large Bohr radius exciton to Coulomb screening is well known and one has utilized the energy shift of 2s exciton with large Bohr radius ( ~ 6nm) as a sensor to monitor metal-insulator Mott transition in $WSe_2$/ $WS_2$ moiré superlattice[7]. (3) The exciton-exciton interaction[12, 35], which includes a dipole-dipole interaction term and an exchange interaction term. The dipole-dipole interaction between intralayer and interlayer excitons is close to zero, as the intralayer exciton has no static electric dipole. Meanwhile, the exciton-exciton exchange interaction relies on wavefunction spatial overlap[31, 46]. Particularly, the intralayer moiré excitons could have sizable spatial overlap with the interlayer moiré excitons when their hole constituents are both in $WSe_2$. The lowest-energy interlayer excitons generated by the pump beam would have their maximum intensity reside around $R_h^X$ site (Figure 4g); For intralayer moiré exciton *I*, its electron and hole both reside around $R_h^h$ site that has small spatial overlap with interlayer excitons (Figure 4g, moiré exciton *I*), so exchange interaction is negligible here. By contrast, moiré exciton *III* has its hole in $R_h^X$ site, occupying the same site with the hole of interlayer exciton (Figure 4g, moiré exciton *I*), where the exchange interaction can arise from the exchange of holes[35]. The exchange interaction between lowest-energy interlayer exciton and intralayer moiré exciton *III* tends to increase their total energy[31, 46], which is one possible reason for the blue shift of moiré exciton *III*. Our observations suggest that the bandgap renormalization effect dominates for moiré exciton *I* and *II*, while exciton-exciton screening and exchange interaction effects may dominate for exciton *III*.

Moiré exciton *I, II* and *III*'s absorption are all suppressed due to state filling of the interlayer excitons, and their decay rate are approximately within the same orders of magnitude, with minor variations (Figure 4f). The moiré exciton *I* (black) shows a faster decay rate than the moiré exciton *III* (blue). To understand the difference in decay rate, we follow the spatial overlap picture of intralayer and interlayer excitons[36, 47]. In our understanding, the spatial overlap between the holes of intralayer and interlayer excitons determines the extent of absorption suppression. The more spatial overlap between the two holes of intralayer and interlayer exctions, the more suppression for the intralayer transition. After the pump excitation, the hot interlayer excitons would thermalize and follow a thermal distribution near the band edge within sub-picosecond[48]. In real space picture, the lowest-energy interlayer exciton has its hole localized in the $R_h^X$ site[47], which barely overlaps with intralayer moiré exciton *I* that has hole in the $R_h^h$ site, but overlap significantly with moiré exciton *III* that has its hole also in $R_h^X$ site. As for the hot interlayer exciton that has extra kinetic energy, it tends to be more delocalized, contributing to overlap with the hole of exciton *I* in $R_h^h$ site. As the assembly of interlayer excitons further cool down, their spatial distribution gradually shrinks to the $R_h^X$ site, leading to a gradual decrease of the spatial overlap between the holes, accompanied by the recovery of moiré resonance *I*, which would account for the difference in decay rate.

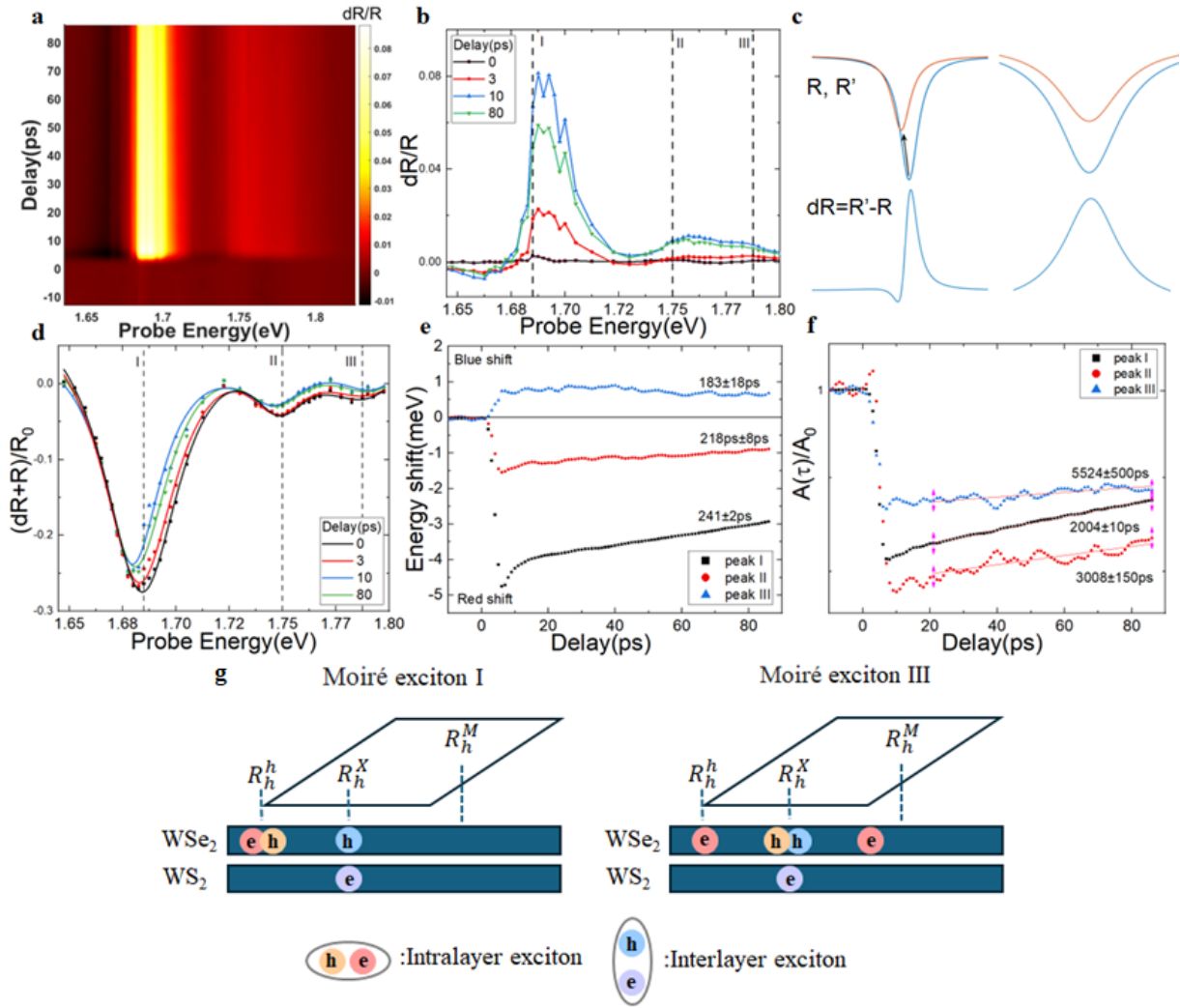

**Figure 4 | Pump probe spectroscopy for WSe₂ intralayer transition**

**a,** Pump-induced reflection change map (*dR/R*) as a function of delay time and the probe energy. **b,** *dR/R* vs Probe energy at *0, 3,10, 80ps*. The moiré excitons *I, II* and *III* are labelled with dash line for reference. **c,** Schematic of the *dR/R* curve as a result of energy shift and suppression of absorption peak. **d,** Dynamics of the reflection contrast, *(dR+R)/R₀*, at various delay times. The Lorentz line shape fitting (colour lines) gives the evolution of three moiré excitons. **e,** Extracted energy red shift as a function of delay time for three moiré excitons. **f,** Extracted absorption amplitude, A(τ)/A₀, as a function of delay time for three moiré excitons. A(τ) is the absorption amplitude at delay τ, A₀ is a constant represents the amplitude before the pump excitation. The vertical axis is in log scale and red straight lines presents single exponential fitting. The lifetimes probed at the resonance energies of three excitons are estimated through single exponential fitting. **g,** Schematic of the Spatial distribution for interlayer exciton and intralayer moiré excitons[36, 47].

## Discussion

We observed a noticeable discrepancy between the dynamics probed at the resonant energies of WS$_2$ and WSe$_2$ intralayer excitons. The lifetime probed at resonant intralayer excitons of WS$_2$ is longer than that of WSe$_2$. As the photo carriers exist in the form of interlayer excitons, the probe at the intralayer exciton energy measures the wavefunction overlap between intralayer excitons (probe) and occupied interlayer excitons. Specifically, the data reflects the wavefunction overlap between the electron (hole) component of intralayer excitons in the WS$_2$ (WSe$_2$) layer and its counterpart in the interlayer excitons. Note that the interlayer exciton has its electron and hole localized around the $R_h^X$ site[47]. The photoluminescence and reflection contrast measurements (Figure 2) indicate that due to the in-plane periodic strain introduced by the lattice reconstruction, the intralayer exciton in WSe$_2$ layer splits into three moiré excitons (*I, II* and *III*). On the other hand, the intralayer exciton in WS$_2$ layer is not much different from that in a pristine layer, spreading uniformly across the moiré cell. The reflection probed at intralayer exciton resonance in WS$_2$ layer well captures the interlayer exciton lifetime. By contrast, the intralayer moiré excitons in WSe$_2$ layer are localized with distinct spatial profiles within a moiré cell [36]. Especially, exciton *I* resides at $R_h^X$ site, away from the lowest-energy interlayer exciton in spatial distribution. For excited interlayer excitons carry extra energies[46], they can have extended distributions and therefore build a sizable spatial overlap with the hole component of the intralayer exciton *I*. As interlayer excitons further cool down and their wave function intensities concentrate around the $R_h^X$ site, the overlap with intralayer excitons *I* gradually shrinks and it displays a fast decay rate. Therefore, the signal probed at intralayer exciton *I* energy primarily reflects the cooling process of interlayer excitons. Intralayer exciton *III*, however, tends to reside at $R_h^X$ site and shares the similar hole component spatial distribution as the lowest-energy interlayer exciton. Therefore the probe at intralayer exciton *III* shows a slower decay rate than that at intralayer exciton *I*. Nonetheless, its decay pattern is still different from that probed at WS$_2$ intralayer exciton. This could be attributed to the weak signal *vs*. noise ratio data at WSe$_2$ intralayer exciton *III* owing to it weak oscillator strength. Or the hole components at interlayer exciton and WSe$_2$ intralayer exciton III are located at different sub-bands. This needs further investigation.

Yet we cannot quantitatively extract the interlayer exciton lifetime in probing at WS$_2$ intralayer exciton energy, for the nearly zero decay pattern in the detection time range. The lower bound that we can determine is approximately from the exciton III's decay pattern. Given exciton III occupies the same site with low-energy interlayer excitons, exciton III's lifetime reflects interlayer exciton lifetime. Therefore, we infer that the interlayer exciton lifetime exceeding *5ns*.

## Conclusion

In this work, we observed that a near-zero twisted WSe$_2$/WS$_2$ heterostructure exhibits spectroscopic features modulated by a signature moiré potential: the ladder-like interlayer photoluminescence and the splitting of moiré intralayer excitons in WSe$_2$ layer. The former reveals the on-site exciton-exciton repulsion between moiré interlayer excitons localized at the same site, while the latter reflects the influence of moiré potential and in-plane strain modification on exciton dispersion. The distinct splitting behaviours in absorption spectra and the slow decay pattern in the pump-probe spectra indicate that the intralayer exciton is delocalized in WS$_2$ layer and the intralayer moiré excitons in WSe$_2$ layer tend to be localized at different sites in the moiré cell, as a result of different in-plane strain and reconstruction. Our pump-probe data confirms that the moiré interlayer exciton experiences a few picoseconds cooling-down time and a lifetime exceeding five nanoseconds. Our study provides a comprehensive approach to investigate the dynamics of moiré interlayer excitons.

**ACKNOWLEDGMENTS**

The work was supported by the National Key R&D Program of China (2020YFA0309600), Guangdong-Hong Kong Joint Laboratory of Quantum Matter and the University Grants Committees/Research Grants Council of Hong Kong SAR (AoE/P-701/20, 17300520, 17301223). K.W. and T.T. acknowledge support from the Elemental Strategy Initiative conducted by the MEXT, Japan (Grant Number JPMXP0112101001) and JSPS KAKENHI (Grant Numbers 19H05790, 20H00354 and 21H05233). H.Y. acknowledges support by NSFC under Grant No. 12274477 and the Department of Science and Technology of Guangdong Province in China (No. 2019QN01X061).

**Author Contributions**

X.D.C and K.F conceived the research. K.F, X.W, Y.Q.X and Y.H.Y fabricated van der Waals heterostructures. K.W. and T.T. grew hexagonal boron nitride crystals. K.F carried out optical measurements. X.D.C, K.F., X.W performed data analysis and interpreted the results. X.D.C, K.F., W.Y and H.Y.Y wrote the paper with input from all authors. All authors discussed the results.

**Conflict of Interest**

The authors have no conflicts to disclose.